\documentclass[fleqn]{2023SCGE}
\usepackage{orcidlink}

\usepackage{multirow}
\newcommand{\nc}{\newcommand*}
\nc{\mU}{{\mathcal{U}}}
\nc{\Msun}{M_\odot}             
\nc{\fpbh}{f_{\mathrm{PBH}}}    
\nc{\be}{\begin{equation}}
\nc{\ee}{\end{equation}}

\nc{\red}[1]{\textcolor{red}{#1}}
\nc{\Eq}[1]{Eq.~\eqref{#1}}     
\nc{\Fig}[1]{Fig.~\ref{#1}}     
\nc{\Table}[1]{Table~\ref{#1}}  
\nc{\Sec}[1]{Sec.~\ref{#1}}     

\begin{document}

\ensubject{subject}

\ArticleType{Article}
\SpecialTopic{SPECIAL TOPIC: }
\Year{2024}
\Month{}
\Vol{}
\No{}
\DOI{}
\ArtNo{}
\ReceiveDate{}
\AcceptDate{}
\OnlineDate{}

\title{Is PSR J0514$-$4002E in a PBH-NS binary?}{Is PSR J0514$-$4002E in a PBH-NS binary?}

\author[1,2]{Zu-Cheng~Chen\orcidlink{0000-0001-7016-9934}}{}
\author[3]{Lang~Liu\orcidlink{0000-0002-0297-9633}}{liulang@bnu.edu.cn}

\AuthorMark{Z.-C. Chen}
\AuthorCitation{Z.-C. Chen, L. Liu}

\address[1]{Department of Physics and Synergetic Innovation Center for Quantum Effects and Applications, Hunan Normal University, Changsha, Hunan 410081, China}
\address[2]{Institute of Interdisciplinary Studies, Hunan Normal University, Changsha, Hunan 410081, China}
\address[3]{Faculty of Arts and Sciences, Beijing Normal University, Zhuhai 519087, China}

\abstract{
Recent pulsar timing observations using MeerKAT of the eccentric binary millisecond pulsar, PSR J0514$-$4002E, have unveiled a companion with a mass in the mass gap, ranging from $2.09\, M_\odot$ to $2.71\, M_\odot$. This challenges conventional astrophysical scenarios for black hole formation. In this paper, we present an alternative explanation: PSR J0514$-$4002E could be in a PBH-NS binary, with the companion potentially being a primordial black hole formed during the early Universe's first-order phase transition. The associated stochastic gravitational-wave background generated during this phase transition can account for the observed signal from the pulsar timing array, and the abundance of primordial black holes is consistent with constraints from LIGO-Virgo-KAGRA.
}

\keywords{PSR J0514$-$4002E, primordial black hole, pulsar timing array, first-order phase transition}
\PACS{04.30.Db, 04.80.Nn, 95.55.Ym}
\maketitle

\begin{multicols}{2}
\section{Introduction}\label{section1}

Neutron stars (NSs) and astrophysical black holes (ABHs) represent two distinct categories of compact objects in the Universe. NSs, born from the gravitational collapse of massive stars, are primarily composed of protons and neutrons. In contrast, ABHs, characterized by their tremendous mass and density, have behavior solely determined by gravity.  
Although specific mass thresholds for these entities remain uncertain, NSs typically have a maximum mass ranging from $2.2\, \Msun$ to $2.5\, \Msun$~\cite{Legred:2021hdx}, while ABHs with masses below $5\, \Msun$ are rarely observed, establishing a ``mass gap" between the most massive NSs and the least massive ABHs~\cite{Ozel:2010su,Fishbach:2020ryj}. The existence of this mass gap provides valuable insights into the formation mechanisms during supernova events for both types of objects. \Authorfootnote

A recent breakthrough within this mass gap arose from pulsar timing observations  using the Karoo Array Telescope (MeerKAT) of the eccentric binary millisecond pulsar PSR J0514$-$4002E in the globular cluster NGC1851~\cite{Barr:2024wwl}. The findings unveiled a total binary mass of $3.887 \pm 0.004\, \Msun$, with the companion falling within the mass gap, possessing a mass between $2.09\, \Msun$ and $2.71\, \Msun$ at a $95\%$ confidence interval~\cite{Barr:2024wwl}. This enigmatic companion challenges conventional astrophysical formation models, being either a very massive NS or a low-mass ABH.

Beyond ABHs, primordial black holes (PBHs) constitute another class of black holes that could populate the Universe. Formed from the gravitational collapse of overdense regions in the early Universe~\cite{Zeldovich:1967lct,Hawking:1971ei,Carr:1974nx}, PBHs exhibit a wide range of initial masses, from Planck relics of $10^{-8}\,\mathrm{kg}$ to thousands of solar masses, depending on the specific model. The study of PBHs holds immense significance, offering unique insights into various astrophysical and cosmological phenomena. PBHs are not only plausible candidates for dark matter (DM)~\cite{Carr:2016drx}, but they can also contribute to gravitational wave (GW) events detected by LIGO-Virgo-KAGRA (LVK)~\cite{Bird:2016dcv,Sasaki:2016jop} and act as seeds for galaxy and supermassive black hole formation~\cite{Bean:2002kx,Kawasaki:2012kn,Nakama:2017xvq,Carr:2018rid}. The exploration of PBHs, from their formation to properties and abundance, provides a valuable avenue for unraveling fundamental aspects of cosmology and contributes significantly to our broader comprehension of the Universe's evolutionary processes.

In this paper, we explore the intriguing possibility that PSR J0514$-$4002E, in a binary with a compact object in the globular cluster NGC 1851, is a PBH-NS binary by calculating the PBH-NS formation rate. Recent results from the NANOGrav~\cite{NANOGrav:2023gor,NANOGrav:2023hde}, PPTA~\cite{Zic:2023gta,Reardon:2023gzh}, EPTA~\cite{EPTA:2023sfo,Antoniadis:2023ott}, and CPTA~\cite{Xu:2023wog} collaborations independently highlight evidence, with varying levels of significance, for a stochastic GW background (SGWB). By employing Bayesian inference on the NANOGrav 15-year data set, we discovered that if the companion to the pulsar is a PBH resulting from a cosmological first-order phase transition (1stOPT), the associated SGWB generated by the 1stOPT aligns compatibly with the recent PTA signal.

\section{PBH-NS formation rate}\label{rate}

The formation of PBH-NS binaries in galaxies occurs primarily through a process known as 2-body scattering, which involves the emission of GWs~\cite{Sasaki:2018dmp,Sasaki:2021iuc}. 
Within galaxies, these binaries result from interactions between two compact objects -- PBHs and NSs -- exchanging energy and momentum, leading to GW emission.

When a PBH approaches a NS within a critical impact parameter on a hyperbolic orbit, the emitted GWs surpass the initial kinetic energy, forming a bound PBH-NS system. This involves capturing the PBH due to the exchange of energy and momentum through GW emission. The capture cross-section, determining the event's probability, is given by~\cite{Mouri:2002mc}
\begin{align}\label{eq:cross_section}
\sigma = 2\pi \left(\frac{85\pi}{6\sqrt{2}}\right)^{2/7}G^2M^{12/7}\mu^{2/7}c^{-10/7}v_\mathrm{rel}^{-18/7},
\end{align}
where $M = m_1 + m_2$ is the total mass, $\mu = m_1 m_2/M^2$ is the reduced mass, $c$ is the speed of light, $v_\mathrm{rel}$ is the relative velocity, and $G$ is the gravitational constant.

To calculate the PBH-NS formation rate, we need to determine the density overlap between the NS and PBH distributions. Adopting the Navarro-Frenk-White (NFW) model~\cite{Navarro:1995iw} for the galactic halo profile within the context of DM, we express the DM density as
\begin{equation} \label{eq:nfw}
\rho_\mathrm{DM}(r) = \rho_0\left[\frac{r}{R_\mathrm{s}}\left(1 + \frac{r}{R_\mathrm{s}}\right)^2\right]^{-1},
\end{equation}
where $\rho_0$ and $R_s$ represent the characteristic density and radius of the halo, respectively.
This model, a reasonable approximation for the DM distribution within galaxies, allows us to focus on general characteristics and trends without delving into intricate aspects.

\begin{figure}[H]
\centering
\includegraphics[width=0.45\textwidth]{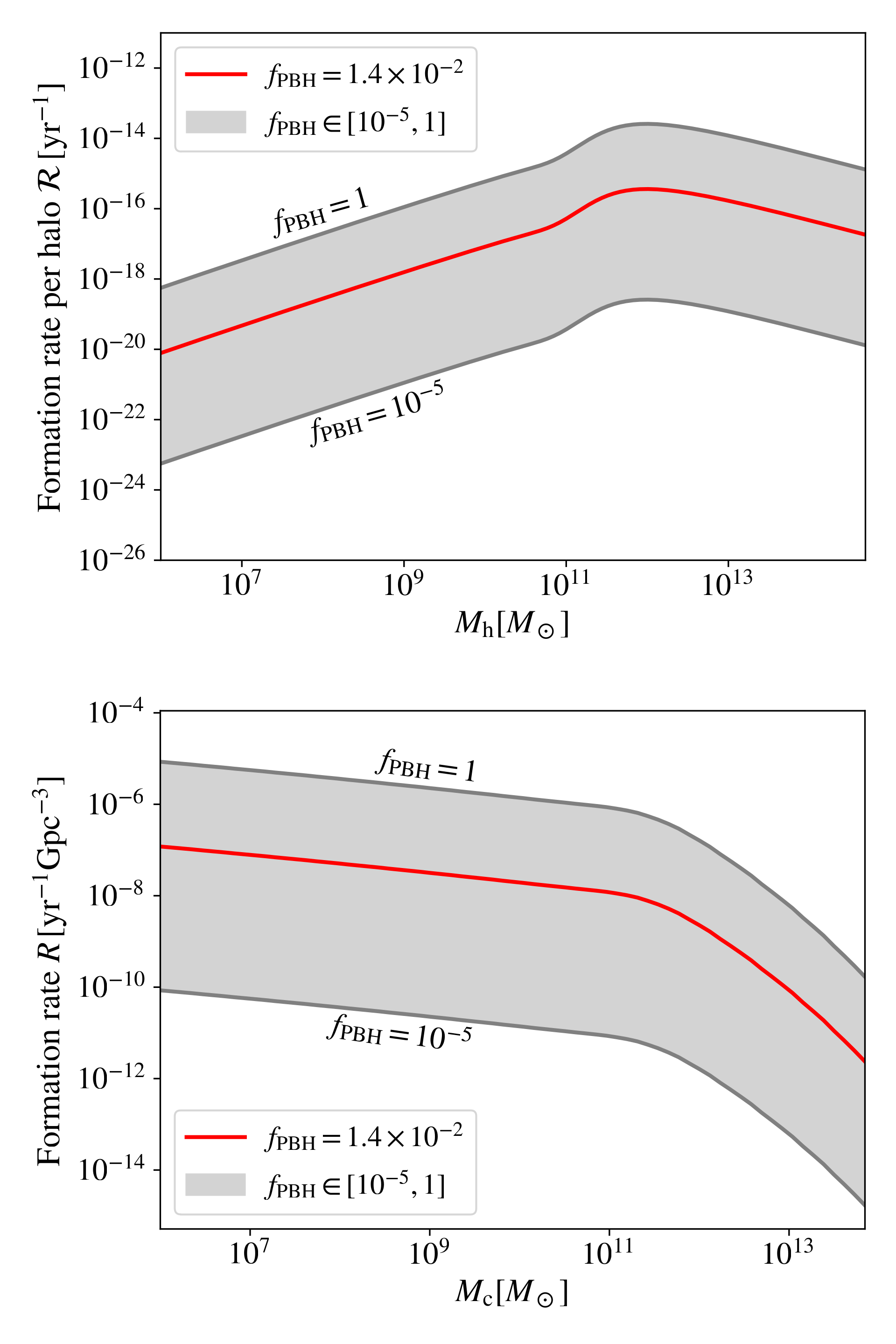}
\caption{\label{fig:MhMc}\textbf{Upper panel}: Formation rate of PBH-NS binaries per halo, $\mathcal{R}$, as a function of halo mass $M_\mathrm{h}$. The red solid curve represents the upper limit, $\fpbh \lesssim 1.4\times 10^{-2}$, derived from the non-detection of SGWB produced by PBH binaries with LVK~\cite{Chen:2019irf}. The shaded gray region encompasses the range of PBH abundance within $\fpbh \in [10^{-5}, 1]$.
\textbf{Lower panel}: Formation rate of PBH-NS integrated over the halo mass, $\mathrm{R}$, as a function of the smallest contributing halo with mass $M_c$.}
\end{figure}

PBHs originate from the early Universe, while NSs form through the gravitational collapse and supernova explosions of massive stars. NSs often acquire substantial ``natal kick" velocities, reaching speeds of hundreds of kilometers per second~\cite{Arzoumanian:2001dv}. In our analysis, we adopt an exponential distribution for the galactic population of NSs, which is well-motivated by the star formation history~\cite{Sartore:2009wn}. Specifically, we employ the following spherically symmetric model
\begin{align} \label{eq:nsdis}
    \rho_\mathrm{NS}(r) = \rho^{0}_\mathrm{NS}e^{-r/R_\mathrm{NS}}~,
\end{align}
where $\rho^{0}_\mathrm{NS}$ represents the characteristic density and $R_\mathrm{NS}$ is the radius associated with the distribution.
While the NS distribution model in \Eq{eq:nsdis} is primarily based on studies of the Milky Way, we acknowledge that different galaxy types may exhibit varying NS distributions. For non-spiral galaxies, the actual distribution could differ significantly. However, since NGC 1851 (the host of PSR J0514$-$4002E) is a globular cluster typically found in galactic halos, our primary conclusions remain robust as they depend more strongly on the local cluster dynamics than the global galactic distribution.
The resulting binary formation rate within a specific galactic halo can be expressed as
\begin{align}\label{eq:rate_integral}
    \mathcal{R} = 4 \pi \int_0^{R_\mathrm{vir}} \mathrm{d} r\, r^2 \frac{\rho_\mathrm{NS}}{m_1}\frac{\rho_\mathrm{PBH}}{m_2}\langle\sigma v_\mathrm{rel}\rangle.
\end{align}
Here, $m_1$ and $m_2$ are the masses of the NS and the PBH, respectively. Furthermore, $R_\mathrm{vir}$ denotes the virial radius of the halo, and $\langle\sigma v_\mathrm{rel}\rangle$ represents the velocity-averaged product of the capture cross-section and the relative velocity between the PBH and the NS.

In~\Eq{eq:rate_integral}, we integrate over the radial distance within the halo, considering contributions from the NS and PBH distributions. The resulting formation rate measures the frequency with which PBH-NS systems form. By considering the density profiles, mass ratios, and velocity distributions, we can gain insights into the formation and prevalence of PBH-NS binaries in different galactic environments. 
To calculate the total formation rates of the PBH-NS binaries in the galaxy, we perform a convolution of the per-halo formation rates with the halo mass function, ${\rm d}n/{\rm d}M_{\mathrm{h}}$. This convolution process is mathematically expressed as
\begin{equation} \label{eq:totrate}
   \mathrm{R} = \int_{M_c} \mathcal{R} \frac{{\rm d}n}{{\rm d}M_\mathrm{h}} {\rm d}M_\mathrm{h}~,
\end{equation}
where $M_{\mathrm{c}}$ represents the lower cutoff limit for the halos that contribute to the formation rates.

In~\Fig{fig:MhMc}, the upper panel shows the PBH-NS binary formation rate per halo, $\mathcal{R}$, versus halo mass $M_\mathrm{h}$. The red solid curve represents the upper limit ($\fpbh \lesssim 1.4\times 10^{-2}$) from the non-detection of SGWB produced by PBH binaries with LVK~\cite{Chen:2019irf}. The lower panel displays the integrated formation rate of PBH-NS on the halo mass, $\mathrm{R}$, versus the smallest contributing halo mass $M_c$.

\section{PBH produced by 1stOPT}\label{PBH}
PBHs may be abundantly produced during a supercooled 1stOPT through a captivating phenomenon known as ``late-blooming"~\cite{Liu:2021svg}. In this process, bubbles form randomly throughout the false vacuum volume during the 1stOPT. As the Universe nears the percolation point, there is a non-zero probability of encountering regions, roughly the size of the Hubble radius, where bubble nucleation has not initiated. These regions, termed ``late-bloomers", maintain constant vacuum energy throughout the supercooled 1stOPT, gradually decreasing, reminiscent of radiation, due to redshifting.

After percolation, these ``late-bloomers" transform into denser regions. If these Hubble-sized regions exceed a critical density threshold, typically $\delta \rho/\rho \gtrsim 0.45$, they undergo gravitational collapse, forming PBHs. In such cases, the PBH mass can be estimated as~\cite{Gouttenoire:2023naa,Gouttenoire:2023bqy}
\begin{equation}
M_{\mathrm{PBH}} \simeq M_{\odot}\left(\frac{20}{g_*(T_{\mathrm{eq}})}\right)^{1 / 2}\left(\frac{0.14\, \mathrm{GeV}}{T_{\mathrm{eq}}}\right)^2.
\end{equation}
We refer to~\cite{Gouttenoire:2023naa} for a comprehensive mathematical treatment and analytical expressions related to PBH formation through the ``late-blooming" mechanism.

The 1stOPTs have a dual impact: they can lead to PBH formation and generate an SGWB. Here, we adopt the bulk flow model for the GW signal~\cite{Jinno:2017fby,Konstandin:2017sat,Lewicki:2020jiv,Lewicki:2020azd,Lewicki:2020azd,Cutting:2020nla}. For relativistic bubble wall velocities, the bulk flow model yields the energy density for GWs as~\cite{Konstandin:2017sat}
\be
\Omega_{\mathrm{GW}} h^2 \simeq \frac{10^{-6}}{\left(g_* / 100\right)^{1 / 3}}\left(\frac{H_*}{\beta}\right)^2\left(\frac{\alpha}{1+\alpha}\right)^2 S_{\mathrm{PT}}(f) S_H(f),
\ee
where the spectral shape $S_{\mathrm{PT}}(f)$ is defined as
\be
S_{\mathrm{PT}}(f)=\frac{3\left(f / f_{\mathrm{PT}}\right)^{0.9}}{2.1+0.9\left(f / f_{\mathrm{PT}}\right)^3}, \quad f_{\mathrm{PT}}=\left(\frac{a_*}{a_0}\right) 0.8\left(\frac{\beta}{2 \pi}\right),
\ee
and the redshift factor between the percolation epoch ``$*$" and today ``0" is given by
\be
a_* / a_0=1.65 \times 10^{-2} \mathrm{mHz}\left(\frac{T_{\mathrm{eq}}}{100 \mathrm{GeV}}\right)\left(\frac{g_{\mathrm{eff}, \mathrm{reh}}}{100}\right)^{1 / 6} H_*^{-1} .
\ee
To refine our model, we introduce a correction factor
\be
S_H(f)=\frac{\left(f / f_*\right)^{2.1}}{1+\left(f / f_*\right)^{2.1}}, \quad f_*=c_*\left(\frac{a_*}{a_0}\right)\left(\frac{H_*}{2 \pi}\right),
\ee
where $c_*$, an $\mathcal{O}(1)$ constant, ensures an $f^3$ scaling for emitted frequencies smaller than the Hubble factor $H_* /(2 \pi)$, following causality~\cite{Durrer:2003ja,Caprini:2009fx,Cai:2019cdl,Hook:2020phx}. In our analysis, we treat $c_*$ as a free parameter that is determined by PTA data.

\section{Data analysis and result}\label{Data}

We conduct a thorough analysis using the NANOGrav 15-year data set~\cite{NANOGrav:2023hde} to constrain the 1stOPT model via Bayesian inference. By employing amplitude values from the free spectrum provided by NANOGrav, considering Hellings-Downs~\cite{Hellings:1983fr} correlations indicative of an SGWB within the framework of general relativity, we utilize observations from $68$ pulsars spanning an observation time of $T_\mathrm{obs}=16.03\,\mathrm{yr}$~\cite{NANOGrav:2023hde}.

\begin{figure}[H]
	\centering
 \includegraphics[width=0.5\textwidth]{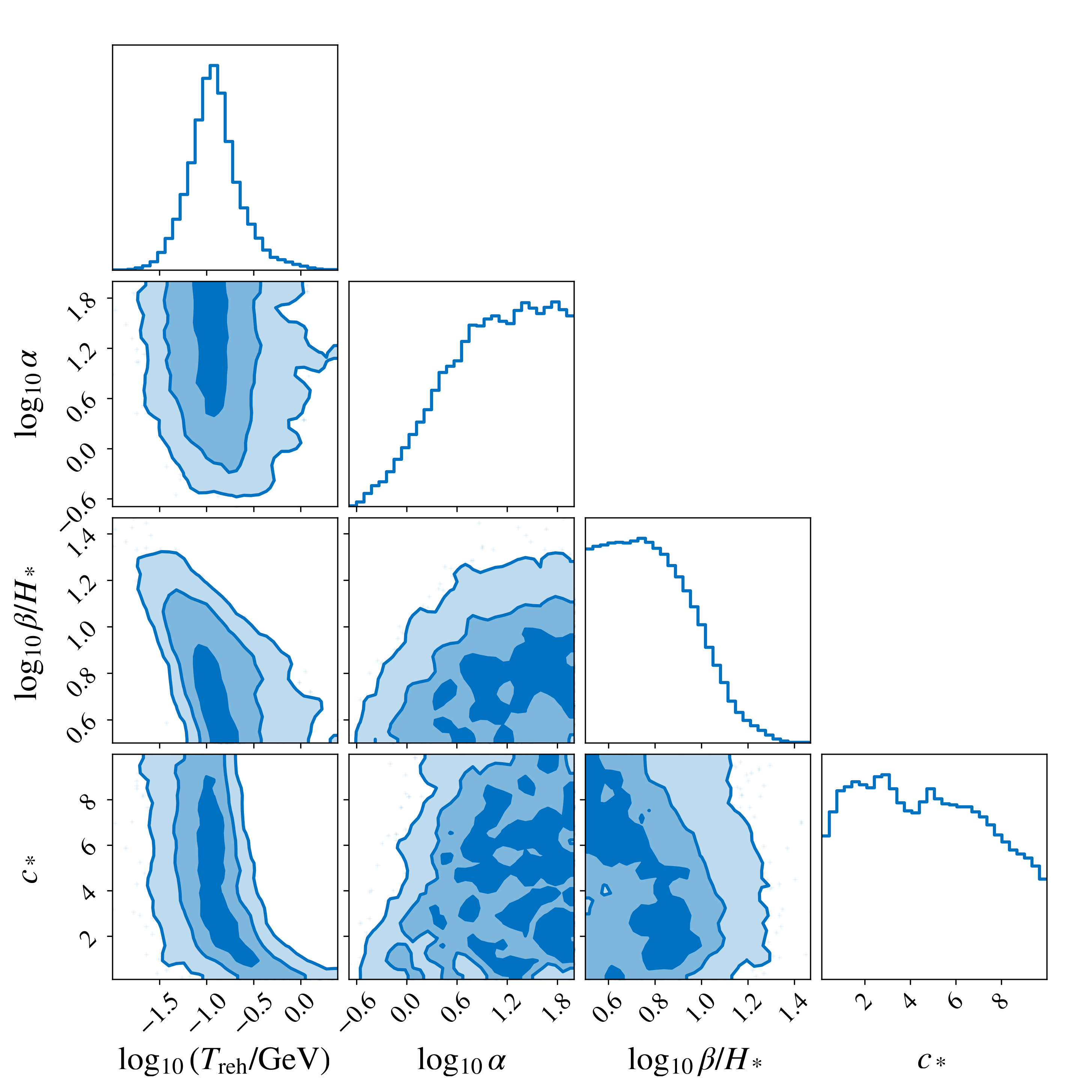}
	\caption{\label{fig:posts_all}Posterior distributions for the model parameters $\Lambda\equiv \{T_\mathrm{reh}, \alpha, \beta/H_*, c_*\}$ obtained from the NANOGrav 15-yr data set. The contours in the two-dimensional plot represent the credible regions at $1 \sigma$, $2 \sigma$, and $3 \sigma$ levels.}
\end{figure}

Starting with the posterior data for the time delay $d(f)$, released by NANOGrav~\cite{NANOGrav:2023gor}, we derive the power spectrum $S(f)$ from this time delay $d(f)$ through
\begin{equation}
S(f) = d(f)^2\, T_{\mathrm{obs}}.
\end{equation}
Subsequently, we compute the GW energy density, $\hat{\Omega}_{\mathrm{GW}}(f)$, of the free spectrum as
\begin{equation}
\label{hatomega}
\hat{\Omega}_{\mathrm{GW}}(f)=\frac{2 \pi^2}{3 H_0^2} f^2 h_c^2(f) = \frac{8\pi^4}{H_0^2} T_{\mathrm{obs}} f^5 d^2(f),
\end{equation}
where the characteristic strain $h_c(f)$ is defined by
\begin{equation}
h_c^2(f)=12 \pi^2 f^3 S(f).
\end{equation}
For each observed frequency $f_i$, we employ the obtained posteriors $\hat{\Omega}_{\mathrm{GW}}(f_i)$ from \Eq{hatomega}, to approximate the corresponding kernel density, $\mathcal{L}_i(\hat{\Omega}_{\mathrm{GW}})$. We then compute the total log-likelihood as the sum of individual log-likelihoods by~\cite{Moore:2021ibq,Lamb:2023jls,Liu:2023ymk}
\begin{equation}
\ln \mathcal{L}(\Lambda) = \sum_{i=1}^{66} \ln \mathcal{L}_i(\Omega_{\mathrm{GW}}(f_i, \Lambda)),
\end{equation}
where $\Lambda\equiv \{T_\mathrm{reh}, \alpha, \beta/H_*, c_*\}$ denotes the set of four model parameters. Our analysis encompasses a spectrum of $14$ frequencies~\cite{NANOGrav:2023gor}, spanning from $1.98$\,nHz to $27.7$\,nHz, and we employ the \texttt{dynesty} sampler~\cite{Speagle:2019ivv} integrated into the \texttt{Bilby} package~\cite{Ashton:2018jfp,Romero-Shaw:2020owr} for exploring the parameter space.

\begin{figure}[H]
	\centering
	\includegraphics[width=0.5\textwidth]{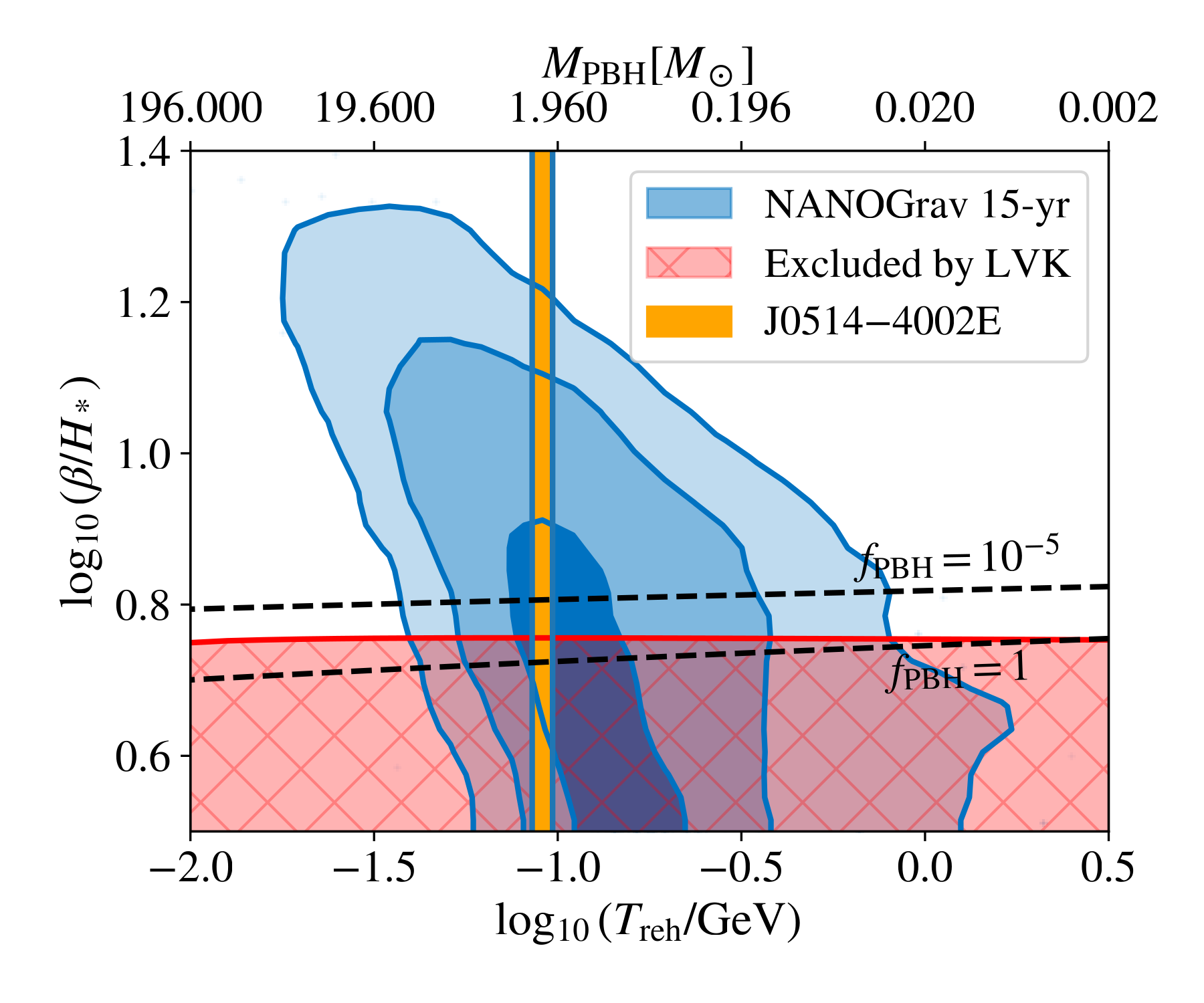}
	\caption{\label{fig:T_beta}Two-dimensional posterior distributions for the $T_\mathrm{reh}$ and $\beta/H_*$ parameters in blue contours, derived from the NANOGrav 15-year data set. The contours represent the $1\sigma$, $2\sigma$, and $3\sigma$ credible intervals, respectively. The red region denotes exclusion by the LVK collaboration, based on the non-detection of the SGWB in their first three observing runs. Two black dashed lines are drawn to mark $\fpbh=1$ and $\fpbh=10^{-5}$, respectively. The orange region highlights the mass range $[2.09\Msun, 2.71\Msun]$ associated with the compact object companion to the pulsar in the PSR J0514$-$4002E system.}
\end{figure}

In our analysis, we use uniform priors for each parameter: $\log_{10} (T_\mathrm{reh}/\mathrm{GeV})$ within $[-3, 1]$, $\log_{10} \alpha$ within $[-2, 2]$, $\log_{10} (\beta/H_*)$ within $[0.5, 3]$, and $c_*$ within $[0.1, 10]$. The posterior distributions for these model parameters are illustrated in~\Fig{fig:posts_all}. Specifically, we find $\log_{10} (T_\mathrm{reh}/\mathrm{GeV}) = -0.93^{+0.50}_{-0.41}$, and $c_* = 4.62^{+4.67}_{-3.95}$. Unless otherwise specified, we report results using the median value and $90\%$ equal-tail credible interval. Furthermore, we obtain $\log_{10} \alpha \gtrsim 0.006$ and $\log_{10} (\beta/H_*) \lesssim 1.1$ at a $95\%$ credible interval.

In~\Fig{fig:T_beta}, we present the posterior distribution for the $T_\mathrm{reh}$ and $\beta/H_*$ parameters, derived from the NANOGrav 15-year data set. The orange region highlights the mass range $[2.09\Msun, 2.71\Msun]$, corresponding to the mass of the compact object companion to the PSR J0514$-$4002E in a binary system. Interestingly, the mass of the compact object companion resides within the $1\sigma$ credible interval. When incorporating additional constraints from the non-detection of the SGWB by LVK during their first three observing runs, the parameter space undergoes further refinement, yet remains within the $1\sigma$ credible interval. In summary, the PBH-NS binary scenario stands out as a compelling and consistent explanation for the observed binary comprising a compact object and PSR J0514$-$4002E.

\section{\label{Con}Summary and discussion}
In light of the recent MeerKAT pulsar timing observations of the eccentric binary millisecond pulsar PSR J0514$-$4002E, a pivotal discovery has emerged, uncovering a companion with a mass ranging between $2.09, M_\odot$ and $2.71, M_\odot$. This observation challenges conventional astrophysical scenarios for the formation of black holes within the specified mass gap.
Here, we present an alternative hypothesis suggesting that this mysterious compact object could be of primordial origin, specifically a PBH formed during the early Universe's 1stOPT. 
The associated SGWB, generated during this 1stOPT, elegantly accounts for the recently detected stochastic signal by the PTA. Furthermore, our proposed scenario seamlessly aligns with observations from LVK, providing a comprehensive and compelling explanation for the observed compact object companion to PSR J0514$-$4002E.

The potential existence of a PBH-NS binary system introduces profound implications for our understanding of compact objects in the Universe. Notably, it provides a fresh perspective on the longstanding mystery of the mass gap, challenging the conventional notion that it results solely from the upper limit of NS masses and the lower limit of black hole masses. The inclusion of a PBH companion in the binary system suggests a hitherto unexplored avenue for the formation of black holes within the mass gap, thereby redefining our understanding of the mass spectrum in compact object binaries. Moreover, the presence of PBHs, formed during the early Universe's 1stOPT, carries extensive ramifications. These PBHs, previously proposed as candidates for DM, align their abundance with constraints derived from GW detections by LVK. If the companion to PSR J0514$-$4002E indeed proves to be a PBH, it would furnish observational evidence substantiating the existence of these intriguing objects and their potential contribution to the DM content of the Universe. This discovery thus not only addresses a longstanding astrophysical enigma but also intertwines with broader cosmological inquiries regarding the nature and cosmic role of PBHs.

While one might anticipate that enhanced scalar perturbations at small scales during inflation could generate scalar-induced GWs~\cite{Ananda:2006af,Baumann:2007zm}, offering a plausible explanation for the PTA signal and triggering the formation of PBHs, relying solely on scalar-induced GWs tends to lead to excessive PBH production~\cite{NANOGrav:2023hvm}, surpassing the expected DM abundance. Although the introduction of primordial non-Gaussianity can alleviate the PBH overproduction issue~\cite{Franciolini:2023pbf,Liu:2023ymk}, it typically results in PBHs with masses too small ($\lesssim 1\, M_\odot$)~\cite{NANOGrav:2023hvm} to effectively explain the observed compact object companion to PSR J0514$-$4002E.

\Acknowledgements{ZCC is supported by the National Natural Science Foundation of China under Grant No.~12405056 and the innovative research group of Hunan Province under Grant No.~2024JJ1006.
LL is supported by the National Natural Science Foundation of China Grant under Grant No.~12433001.}

\InterestConflict{The authors declare that they have no conflict of interest.}
\bibliographystyle{JHEP}
\bibliography{ref}

\providecommand{\href}[2]{#2}\begingroup\raggedright\begin{thebibliography}{10}

\bibitem{Legred:2021hdx}
I.~Legred, K.~Chatziioannou, R.~Essick, S.~Han and P.~Landry, \emph{{Impact of the PSR J0740+6620 radius constraint on the properties of high-density matter}}, \href{https://doi.org/10.1103/PhysRevD.104.063003}{\emph{Phys. Rev. D} {\bfseries 104} (2021) 063003} [\href{https://arxiv.org/abs/2106.05313}{{\ttfamily 2106.05313}}].

\bibitem{Ozel:2010su}
F.~Ozel, D.~Psaltis, R.~Narayan and J.E.~McClintock, \emph{{The Black Hole Mass Distribution in the Galaxy}}, \href{https://doi.org/10.1088/0004-637X/725/2/1918}{\emph{Astrophys. J.} {\bfseries 725} (2010) 1918} [\href{https://arxiv.org/abs/1006.2834}{{\ttfamily 1006.2834}}].

\bibitem{Fishbach:2020ryj}
M.~Fishbach, R.~Essick and D.E.~Holz, \emph{{Does Matter Matter? Using the mass distribution to distinguish neutron stars and black holes}}, \href{https://doi.org/10.3847/2041-8213/aba7b6}{\emph{Astrophys. J. Lett.} {\bfseries 899} (2020) L8} [\href{https://arxiv.org/abs/2006.13178}{{\ttfamily 2006.13178}}].

\bibitem{Barr:2024wwl}
E.D.~Barr et~al., \emph{{A pulsar in a binary with a compact object in the mass gap between neutron stars and black holes}}, \href{https://doi.org/10.1126/science.adg3005}{\emph{Science} {\bfseries 383} (2024) 275} [\href{https://arxiv.org/abs/2401.09872}{{\ttfamily 2401.09872}}].

\bibitem{Zeldovich:1967lct}
Y.B.~Zel'dovich and I.D.~Novikov, \emph{{The Hypothesis of Cores Retarded during Expansion and the Hot Cosmological Model}}, {\emph{Sov. Astron.} {\bfseries 10} (1967) 602}.

\bibitem{Hawking:1971ei}
S.~Hawking, \emph{{Gravitationally collapsed objects of very low mass}}, \href{https://doi.org/10.1093/mnras/152.1.75}{\emph{Mon. Not. Roy. Astron. Soc.} {\bfseries 152} (1971) 75}.

\bibitem{Carr:1974nx}
B.J.~Carr and S.W.~Hawking, \emph{{Black holes in the early Universe}}, \href{https://doi.org/10.1093/mnras/168.2.399}{\emph{Mon. Not. Roy. Astron. Soc.} {\bfseries 168} (1974) 399}.

\bibitem{Carr:2016drx}
B.~Carr, F.~Kuhnel and M.~Sandstad, \emph{{Primordial Black Holes as Dark Matter}}, \href{https://doi.org/10.1103/PhysRevD.94.083504}{\emph{Phys. Rev. D} {\bfseries 94} (2016) 083504} [\href{https://arxiv.org/abs/1607.06077}{{\ttfamily 1607.06077}}].

\bibitem{Bird:2016dcv}
S.~Bird, I.~Cholis, J.B.~Mu\~noz, Y.~Ali-Ha\"\i{}moud, M.~Kamionkowski, E.D.~Kovetz et~al., \emph{{Did LIGO detect dark matter?}}, \href{https://doi.org/10.1103/PhysRevLett.116.201301}{\emph{Phys. Rev. Lett.} {\bfseries 116} (2016) 201301} [\href{https://arxiv.org/abs/1603.00464}{{\ttfamily 1603.00464}}].

\bibitem{Sasaki:2016jop}
M.~Sasaki, T.~Suyama, T.~Tanaka and S.~Yokoyama, \emph{{Primordial Black Hole Scenario for the Gravitational-Wave Event GW150914}}, \href{https://doi.org/10.1103/PhysRevLett.117.061101}{\emph{Phys. Rev. Lett.} {\bfseries 117} (2016) 061101} [\href{https://arxiv.org/abs/1603.08338}{{\ttfamily 1603.08338}}].

\bibitem{Bean:2002kx}
R.~Bean and J.~Magueijo, \emph{{Could supermassive black holes be quintessential primordial black holes?}}, \href{https://doi.org/10.1103/PhysRevD.66.063505}{\emph{Phys. Rev. D} {\bfseries 66} (2002) 063505} [\href{https://arxiv.org/abs/astro-ph/0204486}{{\ttfamily astro-ph/0204486}}].

\bibitem{Kawasaki:2012kn}
M.~Kawasaki, A.~Kusenko and T.T.~Yanagida, \emph{{Primordial seeds of supermassive black holes}}, \href{https://doi.org/10.1016/j.physletb.2012.03.056}{\emph{Phys. Lett. B} {\bfseries 711} (2012) 1} [\href{https://arxiv.org/abs/1202.3848}{{\ttfamily 1202.3848}}].

\bibitem{Nakama:2017xvq}
T.~Nakama, B.~Carr and J.~Silk, \emph{{Limits on primordial black holes from $\mu$ distortions in cosmic microwave background}}, \href{https://doi.org/10.1103/PhysRevD.97.043525}{\emph{Phys. Rev. D} {\bfseries 97} (2018) 043525} [\href{https://arxiv.org/abs/1710.06945}{{\ttfamily 1710.06945}}].

\bibitem{Carr:2018rid}
B.~Carr and J.~Silk, \emph{{Primordial Black Holes as Generators of Cosmic Structures}}, \href{https://doi.org/10.1093/mnras/sty1204}{\emph{Mon. Not. Roy. Astron. Soc.} {\bfseries 478} (2018) 3756} [\href{https://arxiv.org/abs/1801.00672}{{\ttfamily 1801.00672}}].

\bibitem{NANOGrav:2023gor}
{\scshape NANOGrav} collaboration, \emph{{The NANOGrav 15 yr Data Set: Evidence for a Gravitational-wave Background}}, \href{https://doi.org/10.3847/2041-8213/acdac6}{\emph{Astrophys. J. Lett.} {\bfseries 951} (2023) L8} [\href{https://arxiv.org/abs/2306.16213}{{\ttfamily 2306.16213}}].

\bibitem{NANOGrav:2023hde}
{\scshape NANOGrav} collaboration, \emph{{The NANOGrav 15 yr Data Set: Observations and Timing of 68 Millisecond Pulsars}}, \href{https://doi.org/10.3847/2041-8213/acda9a}{\emph{Astrophys. J. Lett.} {\bfseries 951} (2023) L9} [\href{https://arxiv.org/abs/2306.16217}{{\ttfamily 2306.16217}}].

\bibitem{Zic:2023gta}
A.~Zic et~al., \emph{{The Parkes Pulsar Timing Array third data release}}, \href{https://doi.org/10.1017/pasa.2023.36}{\emph{Publ. Astron. Soc. Austral.} {\bfseries 40} (2023) e049} [\href{https://arxiv.org/abs/2306.16230}{{\ttfamily 2306.16230}}].

\bibitem{Reardon:2023gzh}
D.J.~Reardon et~al., \emph{{Search for an Isotropic Gravitational-wave Background with the Parkes Pulsar Timing Array}}, \href{https://doi.org/10.3847/2041-8213/acdd02}{\emph{Astrophys. J. Lett.} {\bfseries 951} (2023) L6} [\href{https://arxiv.org/abs/2306.16215}{{\ttfamily 2306.16215}}].

\bibitem{EPTA:2023sfo}
{\scshape EPTA} collaboration, \emph{{The second data release from the European Pulsar Timing Array - I. The dataset and timing analysis}}, \href{https://doi.org/10.1051/0004-6361/202346841}{\emph{Astron. Astrophys.} {\bfseries 678} (2023) A48} [\href{https://arxiv.org/abs/2306.16224}{{\ttfamily 2306.16224}}].

\bibitem{Antoniadis:2023ott}
{\scshape EPTA, InPTA:} collaboration, \emph{{The second data release from the European Pulsar Timing Array - III. Search for gravitational wave signals}}, \href{https://doi.org/10.1051/0004-6361/202346844}{\emph{Astron. Astrophys.} {\bfseries 678} (2023) A50} [\href{https://arxiv.org/abs/2306.16214}{{\ttfamily 2306.16214}}].

\bibitem{Xu:2023wog}
H.~Xu et~al., \emph{{Searching for the Nano-Hertz Stochastic Gravitational Wave Background with the Chinese Pulsar Timing Array Data Release I}}, \href{https://doi.org/10.1088/1674-4527/acdfa5}{\emph{Res. Astron. Astrophys.} {\bfseries 23} (2023) 075024} [\href{https://arxiv.org/abs/2306.16216}{{\ttfamily 2306.16216}}].

\bibitem{Sasaki:2018dmp}
M.~Sasaki, T.~Suyama, T.~Tanaka and S.~Yokoyama, \emph{{Primordial black holes\textemdash{}perspectives in gravitational wave astronomy}}, \href{https://doi.org/10.1088/1361-6382/aaa7b4}{\emph{Class. Quant. Grav.} {\bfseries 35} (2018) 063001} [\href{https://arxiv.org/abs/1801.05235}{{\ttfamily 1801.05235}}].

\bibitem{Sasaki:2021iuc}
M.~Sasaki, V.~Takhistov, V.~Vardanyan and Y.-l.~Zhang, \emph{{Establishing the Nonprimordial Origin of Black Hole\textendash{}Neutron Star Mergers}}, \href{https://doi.org/10.3847/1538-4357/ac66da}{\emph{Astrophys. J.} {\bfseries 931} (2022) 2} [\href{https://arxiv.org/abs/2110.09509}{{\ttfamily 2110.09509}}].

\bibitem{Mouri:2002mc}
H.~Mouri and Y.~Taniguchi, \emph{{Runaway merging of black holes: analytical constraint on the timescale}}, \href{https://doi.org/10.1086/339472}{\emph{Astrophys. J. Lett.} {\bfseries 566} (2002) L17} [\href{https://arxiv.org/abs/astro-ph/0201102}{{\ttfamily astro-ph/0201102}}].

\bibitem{Navarro:1995iw}
J.F.~Navarro, C.S.~Frenk and S.D.M.~White, \emph{{The Structure of cold dark matter halos}}, \href{https://doi.org/10.1086/177173}{\emph{Astrophys. J.} {\bfseries 462} (1996) 563} [\href{https://arxiv.org/abs/astro-ph/9508025}{{\ttfamily astro-ph/9508025}}].

\bibitem{Chen:2019irf}
Z.-C.~Chen and Q.-G.~Huang, \emph{{Distinguishing Primordial Black Holes from Astrophysical Black Holes by Einstein Telescope and Cosmic Explorer}}, \href{https://doi.org/10.1088/1475-7516/2020/08/039}{\emph{JCAP} {\bfseries 08} (2020) 039} [\href{https://arxiv.org/abs/1904.02396}{{\ttfamily 1904.02396}}].

\bibitem{Arzoumanian:2001dv}
Z.~Arzoumanian, D.F.~Chernoffs and J.M.~Cordes, \emph{{The Velocity distribution of isolated radio pulsars}}, \href{https://doi.org/10.1086/338805}{\emph{Astrophys. J.} {\bfseries 568} (2002) 289} [\href{https://arxiv.org/abs/astro-ph/0106159}{{\ttfamily astro-ph/0106159}}].

\bibitem{Sartore:2009wn}
N.~Sartore, E.~Ripamonti, A.~Treves and R.~Turolla, \emph{{Galactic neutron stars I. Space and velocity distributions in the disk and in the halo}}, \href{https://doi.org/10.1051/0004-6361/200912222}{\emph{Astron. Astrophys.} {\bfseries 510} (2010) A23} [\href{https://arxiv.org/abs/0908.3182}{{\ttfamily 0908.3182}}].

\bibitem{Liu:2021svg}
J.~Liu, L.~Bian, R.-G.~Cai, Z.-K.~Guo and S.-J.~Wang, \emph{{Primordial black hole production during first-order phase transitions}}, \href{https://doi.org/10.1103/PhysRevD.105.L021303}{\emph{Phys. Rev. D} {\bfseries 105} (2022) L021303} [\href{https://arxiv.org/abs/2106.05637}{{\ttfamily 2106.05637}}].

\bibitem{Gouttenoire:2023naa}
Y.~Gouttenoire and T.~Volansky, \emph{{Primordial black holes from supercooled phase transitions}}, \href{https://doi.org/10.1103/PhysRevD.110.043514}{\emph{Phys. Rev. D} {\bfseries 110} (2024) 043514} [\href{https://arxiv.org/abs/2305.04942}{{\ttfamily 2305.04942}}].

\bibitem{Gouttenoire:2023bqy}
Y.~Gouttenoire, \emph{{First-Order Phase Transition Interpretation of Pulsar Timing Array Signal Is Consistent with Solar-Mass Black Holes}}, \href{https://doi.org/10.1103/PhysRevLett.131.171404}{\emph{Phys. Rev. Lett.} {\bfseries 131} (2023) 171404} [\href{https://arxiv.org/abs/2307.04239}{{\ttfamily 2307.04239}}].

\bibitem{Jinno:2017fby}
R.~Jinno and M.~Takimoto, \emph{{Gravitational waves from bubble dynamics: Beyond the Envelope}}, \href{https://doi.org/10.1088/1475-7516/2019/01/060}{\emph{JCAP} {\bfseries 01} (2019) 060} [\href{https://arxiv.org/abs/1707.03111}{{\ttfamily 1707.03111}}].

\bibitem{Konstandin:2017sat}
T.~Konstandin, \emph{{Gravitational radiation from a bulk flow model}}, \href{https://doi.org/10.1088/1475-7516/2018/03/047}{\emph{JCAP} {\bfseries 03} (2018) 047} [\href{https://arxiv.org/abs/1712.06869}{{\ttfamily 1712.06869}}].

\bibitem{Lewicki:2020jiv}
M.~Lewicki and V.~Vaskonen, \emph{{Gravitational wave spectra from strongly supercooled phase transitions}}, \href{https://doi.org/10.1140/epjc/s10052-020-08589-1}{\emph{Eur. Phys. J. C} {\bfseries 80} (2020) 1003} [\href{https://arxiv.org/abs/2007.04967}{{\ttfamily 2007.04967}}].

\bibitem{Lewicki:2020azd}
M.~Lewicki and V.~Vaskonen, \emph{{Gravitational waves from colliding vacuum bubbles in gauge theories}}, \href{https://doi.org/10.1140/epjc/s10052-021-09232-3}{\emph{Eur. Phys. J. C} {\bfseries 81} (2021) 437} [\href{https://arxiv.org/abs/2012.07826}{{\ttfamily 2012.07826}}].

\bibitem{Cutting:2020nla}
D.~Cutting, E.G.~Escartin, M.~Hindmarsh and D.J.~Weir, \emph{{Gravitational waves from vacuum first order phase transitions II: from thin to thick walls}}, \href{https://doi.org/10.1103/PhysRevD.103.023531}{\emph{Phys. Rev. D} {\bfseries 103} (2021) 023531} [\href{https://arxiv.org/abs/2005.13537}{{\ttfamily 2005.13537}}].

\bibitem{Durrer:2003ja}
R.~Durrer and C.~Caprini, \emph{{Primordial magnetic fields and causality}}, \href{https://doi.org/10.1088/1475-7516/2003/11/010}{\emph{JCAP} {\bfseries 11} (2003) 010} [\href{https://arxiv.org/abs/astro-ph/0305059}{{\ttfamily astro-ph/0305059}}].

\bibitem{Caprini:2009fx}
C.~Caprini, R.~Durrer, T.~Konstandin and G.~Servant, \emph{{General Properties of the Gravitational Wave Spectrum from Phase Transitions}}, \href{https://doi.org/10.1103/PhysRevD.79.083519}{\emph{Phys. Rev. D} {\bfseries 79} (2009) 083519} [\href{https://arxiv.org/abs/0901.1661}{{\ttfamily 0901.1661}}].

\bibitem{Cai:2019cdl}
R.-G.~Cai, S.~Pi and M.~Sasaki, \emph{{Universal infrared scaling of gravitational wave background spectra}}, \href{https://doi.org/10.1103/PhysRevD.102.083528}{\emph{Phys. Rev. D} {\bfseries 102} (2020) 083528} [\href{https://arxiv.org/abs/1909.13728}{{\ttfamily 1909.13728}}].

\bibitem{Hook:2020phx}
A.~Hook, G.~Marques-Tavares and D.~Racco, \emph{{Causal gravitational waves as a probe of free streaming particles and the expansion of the Universe}}, \href{https://doi.org/10.1007/JHEP02(2021)117}{\emph{JHEP} {\bfseries 02} (2021) 117} [\href{https://arxiv.org/abs/2010.03568}{{\ttfamily 2010.03568}}].

\bibitem{Hellings:1983fr}
R.w.~Hellings and G.s.~Downs, \emph{{UPPER LIMITS ON THE ISOTROPIC GRAVITATIONAL RADIATION BACKGROUND FROM PULSAR TIMING ANALYSIS}}, \href{https://doi.org/10.1086/183954}{\emph{Astrophys. J. Lett.} {\bfseries 265} (1983) L39}.

\bibitem{Moore:2021ibq}
C.J.~Moore and A.~Vecchio, \emph{{Ultra-low-frequency gravitational waves from cosmological and astrophysical processes}}, \href{https://doi.org/10.1038/s41550-021-01489-8}{\emph{Nature Astron.} {\bfseries 5} (2021) 1268} [\href{https://arxiv.org/abs/2104.15130}{{\ttfamily 2104.15130}}].

\bibitem{Lamb:2023jls}
W.G.~Lamb, S.R.~Taylor and R.~van Haasteren, \emph{{Rapid refitting techniques for Bayesian spectral characterization of the gravitational wave background using pulsar timing arrays}}, \href{https://doi.org/10.1103/PhysRevD.108.103019}{\emph{Phys. Rev. D} {\bfseries 108} (2023) 103019} [\href{https://arxiv.org/abs/2303.15442}{{\ttfamily 2303.15442}}].

\bibitem{Liu:2023ymk}
L.~Liu, Z.-C.~Chen and Q.-G.~Huang, \emph{{Implications for the non-Gaussianity of curvature perturbation from pulsar timing arrays}}, \href{https://doi.org/10.1103/PhysRevD.109.L061301}{\emph{Phys. Rev. D} {\bfseries 109} (2024) L061301} [\href{https://arxiv.org/abs/2307.01102}{{\ttfamily 2307.01102}}].

\bibitem{Speagle:2019ivv}
J.S.~Speagle, \emph{{dynesty: a dynamic nested sampling package for estimating Bayesian posteriors and evidences}}, \href{https://doi.org/10.1093/mnras/staa278}{\emph{Mon. Not. Roy. Astron. Soc.} {\bfseries 493} (2020) 3132} [\href{https://arxiv.org/abs/1904.02180}{{\ttfamily 1904.02180}}].

\bibitem{Ashton:2018jfp}
G.~Ashton et~al., \emph{{BILBY: A user-friendly Bayesian inference library for gravitational-wave astronomy}}, \href{https://doi.org/10.3847/1538-4365/ab06fc}{\emph{Astrophys. J. Suppl.} {\bfseries 241} (2019) 27} [\href{https://arxiv.org/abs/1811.02042}{{\ttfamily 1811.02042}}].

\bibitem{Romero-Shaw:2020owr}
I.M.~Romero-Shaw et~al., \emph{{Bayesian inference for compact binary coalescences with bilby: validation and application to the first LIGO\textendash{}Virgo gravitational-wave transient catalogue}}, \href{https://doi.org/10.1093/mnras/staa2850}{\emph{Mon. Not. Roy. Astron. Soc.} {\bfseries 499} (2020) 3295} [\href{https://arxiv.org/abs/2006.00714}{{\ttfamily 2006.00714}}].

\bibitem{Ananda:2006af}
K.N.~Ananda, C.~Clarkson and D.~Wands, \emph{{The Cosmological gravitational wave background from primordial density perturbations}}, \href{https://doi.org/10.1103/PhysRevD.75.123518}{\emph{Phys. Rev. D} {\bfseries 75} (2007) 123518} [\href{https://arxiv.org/abs/gr-qc/0612013}{{\ttfamily gr-qc/0612013}}].

\bibitem{Baumann:2007zm}
D.~Baumann, P.J.~Steinhardt, K.~Takahashi and K.~Ichiki, \emph{{Gravitational Wave Spectrum Induced by Primordial Scalar Perturbations}}, \href{https://doi.org/10.1103/PhysRevD.76.084019}{\emph{Phys. Rev. D} {\bfseries 76} (2007) 084019} [\href{https://arxiv.org/abs/hep-th/0703290}{{\ttfamily hep-th/0703290}}].

\bibitem{NANOGrav:2023hvm}
{\scshape NANOGrav} collaboration, \emph{{The NANOGrav 15 yr Data Set: Search for Signals from New Physics}}, \href{https://doi.org/10.3847/2041-8213/acdc91}{\emph{Astrophys. J. Lett.} {\bfseries 951} (2023) L11} [\href{https://arxiv.org/abs/2306.16219}{{\ttfamily 2306.16219}}].

\bibitem{Franciolini:2023pbf}
G.~Franciolini, A.~Iovino, Junior., V.~Vaskonen and H.~Veermae, \emph{{Recent Gravitational Wave Observation by Pulsar Timing Arrays and Primordial Black Holes: The Importance of Non-Gaussianities}}, \href{https://doi.org/10.1103/PhysRevLett.131.201401}{\emph{Phys. Rev. Lett.} {\bfseries 131} (2023) 201401} [\href{https://arxiv.org/abs/2306.17149}{{\ttfamily 2306.17149}}].

\end{thebibliography}\endgroup
\end{multicols}
\end{document}